\documentclass[aps,twocolumn,superscriptaddress]{revtex4-1}

\usepackage{breqn}
\usepackage{graphicx}
\usepackage{dcolumn}   
\usepackage{bm}        
\usepackage{amssymb}   
\usepackage{soul}
\usepackage{gensymb}
\usepackage{enumerate}
\usepackage{multirow}

\usepackage[breaklinks=true,colorlinks=true,linkcolor=blue,urlcolor=blue,citecolor=blue]{hyperref}

\usepackage{float}

\begin{document}

\title{Assessing the Impact of Alpha Particles on Thermal Confinement in JET D-T Plasmas through Global GENE-Tango Simulations}

\author{A.~Di Siena} 
\affiliation{Max Planck Institute for Plasma Physics Boltzmannstr 2 85748 Garching Germany}
\author{J.~Garcia} 
\affiliation{CEA IRFM F-13108 Saint-Paul-lez-Durance France}
\author{R.~Bilato} 
\affiliation{Max Planck Institute for Plasma Physics Boltzmannstr 2 85748 Garching Germany}
\author{K.~Kirov} 
\affiliation{United Kingdom Atomic Energy Authority Culham Science Centre Abingdon Oxon OX14 3DB United Kingdom of Great Britain and Northern Ireland}
\author{J.~Varela} 
\affiliation{University of Texas Austin Texas USA}
\author{A.~Ba\~n\'on~Navarro}
\affiliation{Max Planck Institute for Plasma Physics Boltzmannstr 2 85748 Garching Germany}
\author{Hyun-Tae~Kim} 
\affiliation{United Kingdom Atomic Energy Authority Culham Science Centre Abingdon Oxon OX14 3DB United Kingdom of Great Britain and Northern Ireland}
\author{C.~Challis} 
\affiliation{United Kingdom Atomic Energy Authority Culham Science Centre Abingdon Oxon OX14 3DB United Kingdom of Great Britain and Northern Ireland}
\author{J.~Hobirk} 
\affiliation{Max Planck Institute for Plasma Physics Boltzmannstr 2 85748 Garching Germany}
\author{A.~Kappatou} 
\affiliation{Max Planck Institute for Plasma Physics Boltzmannstr 2 85748 Garching Germany}
\author{E.~Lerche} 
\affiliation{United Kingdom Atomic Energy Authority Culham Science Centre Abingdon Oxon OX14 3DB United Kingdom of Great Britain and Northern Ireland}
\author{D.~Spong} 
\affiliation{Oak Ridge National Laboratory Oak Ridge Tennessee 37831-8071 US}
\author{C.~Angioni} 
\affiliation{Max Planck Institute for Plasma Physics Boltzmannstr 2 85748 Garching Germany}
\author{T.~G\"orler}
\affiliation{Max Planck Institute for Plasma Physics Boltzmannstr 2 85748 Garching Germany}
\author{E.~Poli} 
\affiliation{Max Planck Institute for Plasma Physics Boltzmannstr 2 85748 Garching Germany}
\author{M.~Bergmann} 
\affiliation{Max Planck Institute for Plasma Physics Boltzmannstr 2 85748 Garching Germany}
\author{F.~Jenko} 
\affiliation{Max Planck Institute for Plasma Physics Boltzmannstr 2 85748 Garching Germany}
\author{JET contributors}
\affiliation{See the author list of “Overview of T and D-T results in JET with ITER-like wall” by CF Maggi et al. to be published in Nuclear Fusion Special Issue: Overview and Summary Papers from the 29th Fusion Energy Conference (London UK 16-21 October 2023)}

\begin{abstract}

The capability of the global, electromagnetic gyrokinetic GENE code interfaced with the transport Tango solver is exploited to address the impact of fusion alpha particles (in their dual role of fast particles and heating source) on plasma profiles and performance at JET in the discharges with the highest quasi-stationary peak fusion power during the DTE2 experimental campaigns. Employing radially global nonlinear electromagnetic GENE-Tango simulations, we compare results with/without alpha particles and alpha heating. Our findings reveal that alpha particles have a negligible impact on turbulent transport, with GENE-Tango converging to similar plasma profiles regardless of their inclusion as a kinetic species in GENE. On the other hand, alpha heating is found to contribute to the peaking of the electron temperature profiles, leading to a 1keV drop on the on-axis electron temperature when alpha heating is neglected in Tango. The minimal impact of alpha particles on turbulent transport in this JET discharge - despite this being the shot with the highest fusion output - is attributed to the low content of fusion alpha in this discharge. To assess the potential impact of alpha particles on turbulent transport in regimes with higher alpha particle density, as expected in ITER and fusion reactors, we artificially increased the alpha particle concentration to levels expected for ITER. By performing global nonlinear GENE standalone simulations, we found that increasing the alpha particle density beyond five times the nominal value lead to significant overall turbulence destabilization. These results demonstrate that an increased alpha particle concentration can significantly impact transport properties under simulated JET experimental conditions. However, these findings cannot be directly extrapolated to ITER due to the substantial differences in parameters such as plasma size, magnetic field, plasma current, and thermal pressure.

\end{abstract}

\pacs{52.65.y,52.35.Mw,52.35.Ra}

\maketitle


\section{Introduction}

Accurate deuterium and tritium (D–T) fusion power prediction is crucial for optimizing D–T operation scenarios and designing future fusion reactors. The complexity arises from the sensitivity of fusion rate to plasma profiles, which not only directly affect turbulent transport but also influence the fusion rate. The latter, then, impacts the alpha heating and the effects of fusion alphas on turbulence. Contrary to present devices with dominant auxiliary heating, in fusion reactors the main heating source is collisional electron heating by fusion alphas, with partial ion heating via collisional energy exchange. The sensitivity of fusion power to plasma parameters leads to a complex nonlinear nature, where changes in thermal profiles not only affect turbulent transport but also impact heating profiles via modifications of alpha heating, collisional energy exchange and radiative power.

This presents a notable challenge for fusion reactors, where these heating sources are the dominant ones, in contrast to current experiments where most of the heating power is often injected by external heating sources. Addressing this intricate dynamic requires a self-consistent numerical treatment, where turbulence, profiles, heat, and particle sources are evolved and computed self-consistently. This is typically performed by running integrated modelling tools using quasi-linear reduced models to compute the turbulent levels associated with specific plasma profiles, such as Trapped Gyro Landau Fluid (TGLF) \cite{Staebler_PoP_2007,Staebler_PoP_2016} and QuaLiKiz \cite{Bourdelle_PoP_2007,Bourdelle_PPCF_2015,Citrin_PoP_2012,Citrin_PPCF_2017}. However, the improved performance of high-fidelity gyrokinetic codes achieved in recent years (via e.g., GPU porting \cite{Germaschewski_PoP_2021}) enables using these advanced models, rather than the quasi-linear codes, allowing accurate computation of plasma profiles in current devices and more reliable prediction of future fusion reactors.

This effort has led to the coupling of flux-tube gyrokinetic codes with transport models, such as GS2-Trinity, GENE-Trinity \cite{Barnes_PoP_2010}, and CGYRO with a nonlinear optimizer leveraging Gaussian process regression techniques \cite{Rodriguez_Fernandez_2022,Rodriguez_arXiv_2023}. In addition, global gyrokinetic codes have also been coupled with transport solvers like the global version of GENE \cite{Jenko_PoP2000,Goerler_JCP2011} and the transport solver Tango \cite{Shestakov_JCP_2003, Parker_NF_2018, DiSiena_NF_2022, Disiena_NF}, thus including radially global effects crucial for modelling turbulent transport in various regimes, such as profile shearing \cite{Garbet_PoP_1996,Waltz_PoP_2005,Hornsby_NF_2018}, turbulent avalanches \cite{Candy_PRL_2003,Sarazin_PoP_2000}, internal transport barriers \cite{Strugarek_PPCF_2013,DiSiena_PRL_2021,Di_Siena_PPCF_2022}, turbulence spreading \cite{Hahm_PPCF_2004} and energetic particles \cite{Chen_RMP_2016, DiSiena_NF_2023_2}.

To ensure reliable predictions from these high-fidelity codes it is essential to validate their results with experimental data, so that improvements in the prediction models can be identified where necessary. In recent years, a substantial effort has been undertaken to validate these models across various plasma discharges at ASDEX Upgrade \cite{DiSiena_NF_2022,Disiena_NF}, JET \cite{Rodriguez_arXiv_2023, Howard_PoP_2023}, and DIII-D \cite{Rodriguez_arXiv_2023, Howard_PoP_2023} both in L-mode and H-mode plasmas. While these largely successful exercises provide confidence in the reliability of these tools, it is important, particularly for applying these codes in predicting performance in future fusion reactors, to extend the validation to experimental conditions with the highest fusion power achievable in current experiments. The JET D-T experimental campaign DTE2 presents a unique opportunity to validate the current D–T fusion power prediction capabilities ahead of ITER D–T experiments. It is especially significant to assess the impact of fusion-born alpha particles on plasma confinement under conditions relevant to future fusion reactors.

The impact of alpha particles on plasma turbulence is still a largely unexplored topic, so far mostly limited to local gyrokinetic simulations \cite{Garcia_PoP_2018}, despite its critical significance for plasma reactors. This is due to the lack of numerical tools able to retain and capture all the relevant physical effects in a self-consistent manner and with a reasonable computational time. Addressing this requires radially global simulations with an appropriate treatment of velocity space, the evolution of physical sources, plasma profiles, and the inclusion of drift-wave turbulence and alpha particles, making it a multi-scale problem.

Despite its inherent complexity, it is important to investigate how alpha particles can impact drift-wave turbulent transport, similar to what is observed for lower-energy fast particles generated from external heating systems \cite{Garcia_PPCF_2022, Citrin_PPCF_2023}. In particular, low to medium-energy energetic particles have shown the capability to effectively regulate plasma turbulence and induce strong turbulence suppression, particularly on ion-scale turbulence when appropriately optimized \cite{Citrin_PRL_2013,DiSiena_NF_2018,DiSiena_NF_2019}. This holds significant promise for conditions relevant to reactor operation especially during the ramp-up/down phases. However, large confinement degradation has also been observed in both flux-tube and global gyrokinetic simulations in scenarios characterized by unstable energetic particle-driven modes \cite{Citrin_PPCF_2015,DiSiena_JPP_2021,Ishizawa_NF_2021,Biancalani_PPCF_2021,DiSiena_NF_2023_2}. Therefore, extrapolating the effects observed for low to medium-energy particles to alpha particles is particularly complex due to the large differences in energy, phase-space characteristics of their background distributions and the delicate balance between turbulence suppression and turbulence destabilization.

In this paper, we investigate the impact of alpha particles and alpha heating on plasma profiles and performance at JET studying the plasma discharge $\#99912$ achieving the highest quasi-stationary peak fusion power during the DTE2 experimental campaign \cite{Hobirk_NF_2023} in 50-50 D-T. Our analysis was done by performing several radially global GENE-Tango simulations using three different numerical setups: i) retaining alpha heating in Tango while neglecting alpha particles in GENE as a kinetic species; ii) neglecting both alpha heating in Tango and alpha particles in GENE; and iii) retaining both alpha heating in Tango and alpha particles in GENE. The profiles obtained by running GENE-Tango for each of these cases are compared to the experimental measurements, revealing that alpha particles do not directly affect turbulent transport in GENE. However, we observe a notable impact of alpha heating on electron temperature profiles, showing an approximate 1 keV reduction when alpha heating is removed from Tango. To explore the potential influence of alpha particles in other regimes with higher alpha particle concentration, we perform standalone GENE simulations, varying the alpha particle density up to one expected for ITER.

This paper is organized as follows. Section \ref{sec1} provides a detailed description of the JET discharge under examination in this study \cite{Hobirk_NF_2023}. The GENE-Tango coupling is detailed in Section \ref{sec2}, while Section \ref{sec3} outlines the numerical setup used for the GENE-Tango simulations. In Section \ref{sec4}, we present the steady-state plasma profiles derived from various GENE-Tango simulations that either retain or neglect alpha particles in GENE and alpha heating in Tango. A comparison of the plasma profiles computed by GENE-Tango and TGLF-ASTRA is presented in Section \ref{sec5}. Given the minimal effect of alpha particles on turbulent transport observed for this JET discharge, Section \ref{sec6} reports the outcomes of a study involving a scan over the nominal alpha particle density, increased by constant factors, while leaving all the other plasma parameters unchanged. Stability analyses are performed on the final steady-state GENE-Tango profiles in Section \ref{sec7}, and nonlinear global GENE simulations are discussed in Section \ref{sec8}. Conclusions are drawn in Section \ref{sec9}. In addition, the Appendix \ref{sec10} presents a study justifying the use of a single thermal ion species for D-T with an effective mass, representing the mixture of deuterium and tritium, instead of two separate ion species.

\section{Experimental scenario: JET $\#99912$} \label{sec1}

\begin{figure}
\begin{center}
\includegraphics[scale=0.23]{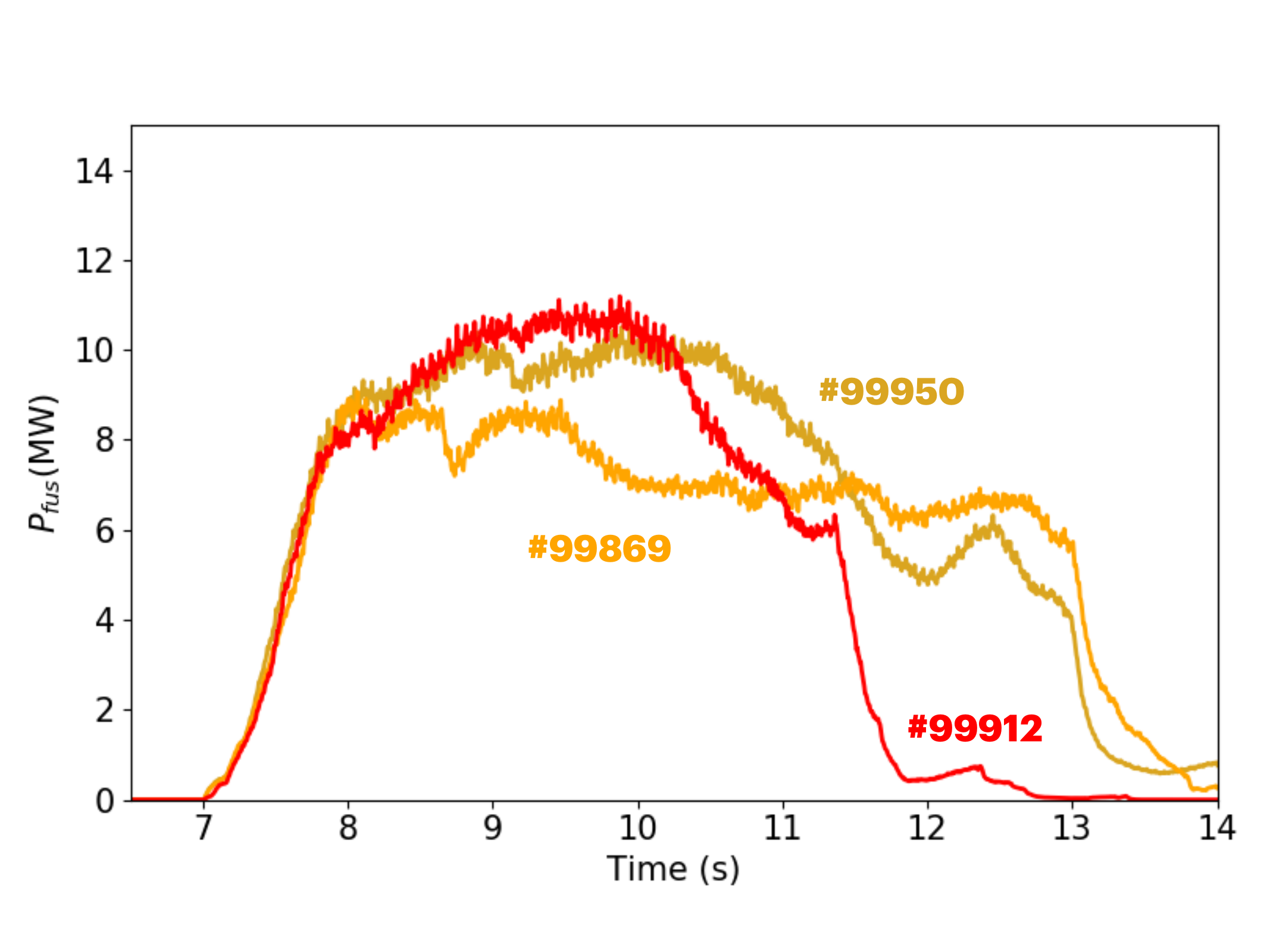}
\par\end{center}
\caption{Time traces of the fusion power ($MW$) for the high-performance JET discharges $\#99869$, $\#99950$, and $\#99912$ during D-T operations, showcasing the best fusion performances in the DTE2 experimental campaign. This figure is adapted from Ref.~\cite{Hobirk_NF_2023}.}
\label{fig:fig_exp_time}
\end{figure}

In this study we simulate the JET discharge $\#99912$ \cite{Hobirk_NF_2023}, characterized by the highest peak fusion output over a quasi-stationary time of $3\tau_{sd,\alpha}$ (with $\tau_{sd,\alpha}$ the alpha particle slowing-down time \cite{Estrada_PoP_2006}) during the entire JET DTE2 experimental campaigns with roughly 50-50 mixture of deuterium and tritium. This significant milestone is illustrated in Fig.~\ref{fig:fig_exp_time}, where the temporal evolution of the fusion output of relevant JET shots is compared. Notably, the figure illustrates that the fusion power of this specific JET discharge exceeds $P_{fus} \approx 10MW$ for a time span exceeding three times the alpha-particle slowing time $\tau_{sd,\alpha}$.

The JET discharge $\#99912$ is an hybrid scenario pulse with deuterium and tritium as main thermal ion species with a nominal on-axis magnetic field strength of approximately $B_0 = 3.45$T. The plasma current is $I_p \approx 2.3$MA, and the safety factor at the flux surface containing 95$\%$ of the poloidal flux is $q_{95} \approx 5$. The auxiliary heating power is comprised of $P_{NBI} \approx 30$MW of Neutral Beam Injection (NBI) and $P_{ICRH} \approx 4$MW of Ion Cyclotron Resonance Heating (ICRH) on hyrogen minority. 
\begin{figure}
\begin{center}
\includegraphics[scale=0.35]{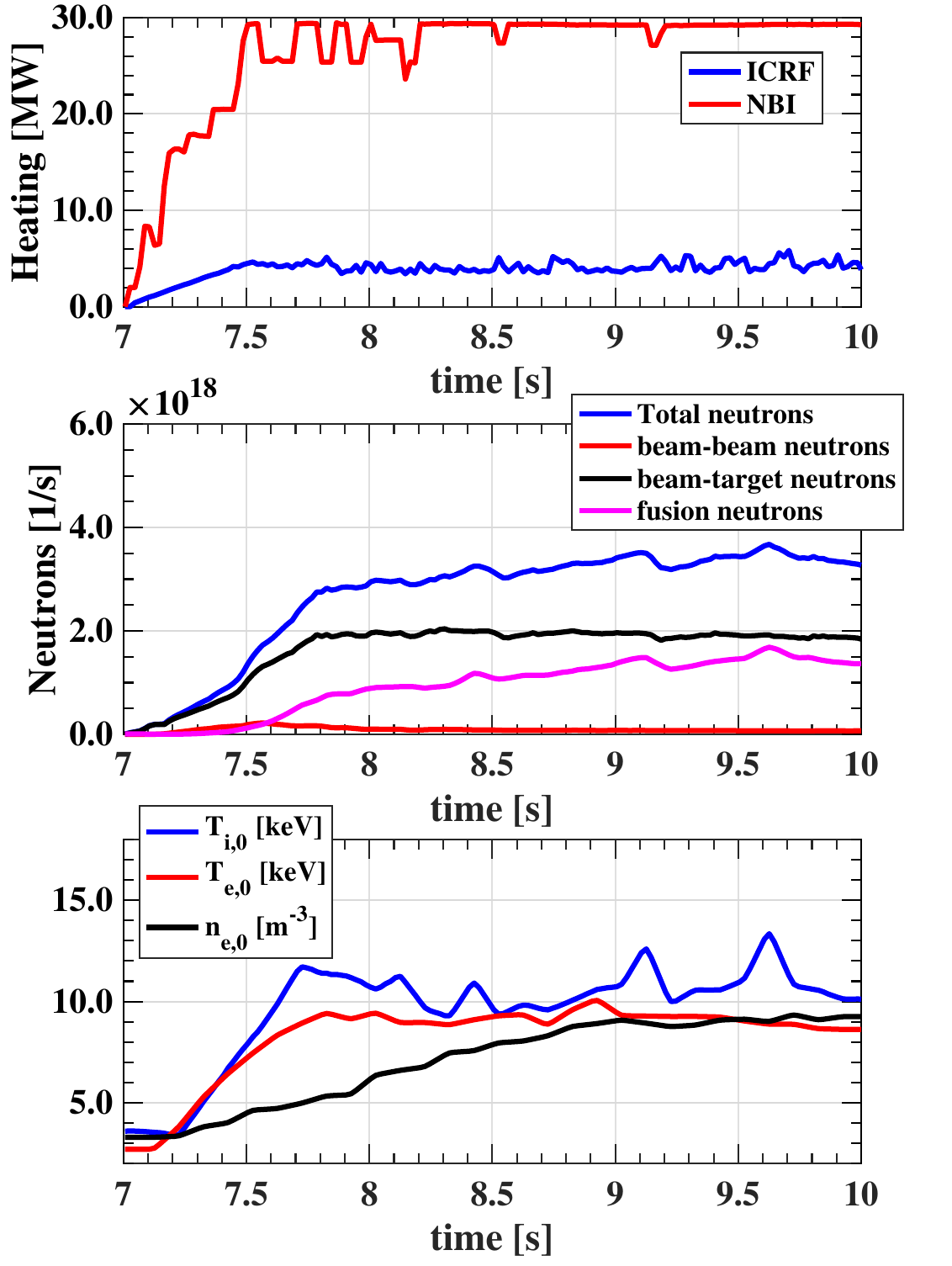}
\par\end{center}
\caption{Time evolution of the (a) ICRH and NBI heating power; (b) total neutron count for beam-beam, beam-target and fusion reactions; (c) on-axis ion, electron temperatures and plasma density for the JET discharge $\#99912$ computed by TRANSP.}
\label{fig:fig_exp}
\end{figure}
The time evolution of the heating powers is illustrated in Fig.~\ref{fig:fig_exp} together with the time-traces of the neutrons and on-axis plasma temperatures and density. Notably, Fig.~\ref{fig:fig_exp}b shows that a large fraction of neutrons originates from beam-target reactions (approximately $55\%$). Fusion-born neutrons, at their peak, constitute only $\sim 45\%$ of the total neutron count, while the contribution of neutrons from beam-beam reactions remains negligible throughout the entire plasma discharge \cite{Kirov_NF_2021}. A more detailed description of this plasma discharge can be found in Ref.~\cite{Hobirk_NF_2023}. The numerical simulations outlined in this paper are performed on the time slice at $t = 8.6$s, occurring within the period when the neutron count reaches the largest value steadily over several alpha-particle slowing down time. In this specific time interval, the alpha particle density reaches its maximum concentration, indicating the time where alpha particle effects on plasma profiles and turbulence are expected to be more pronounced.
\begin{figure}
\begin{center}
\includegraphics[scale=0.30]{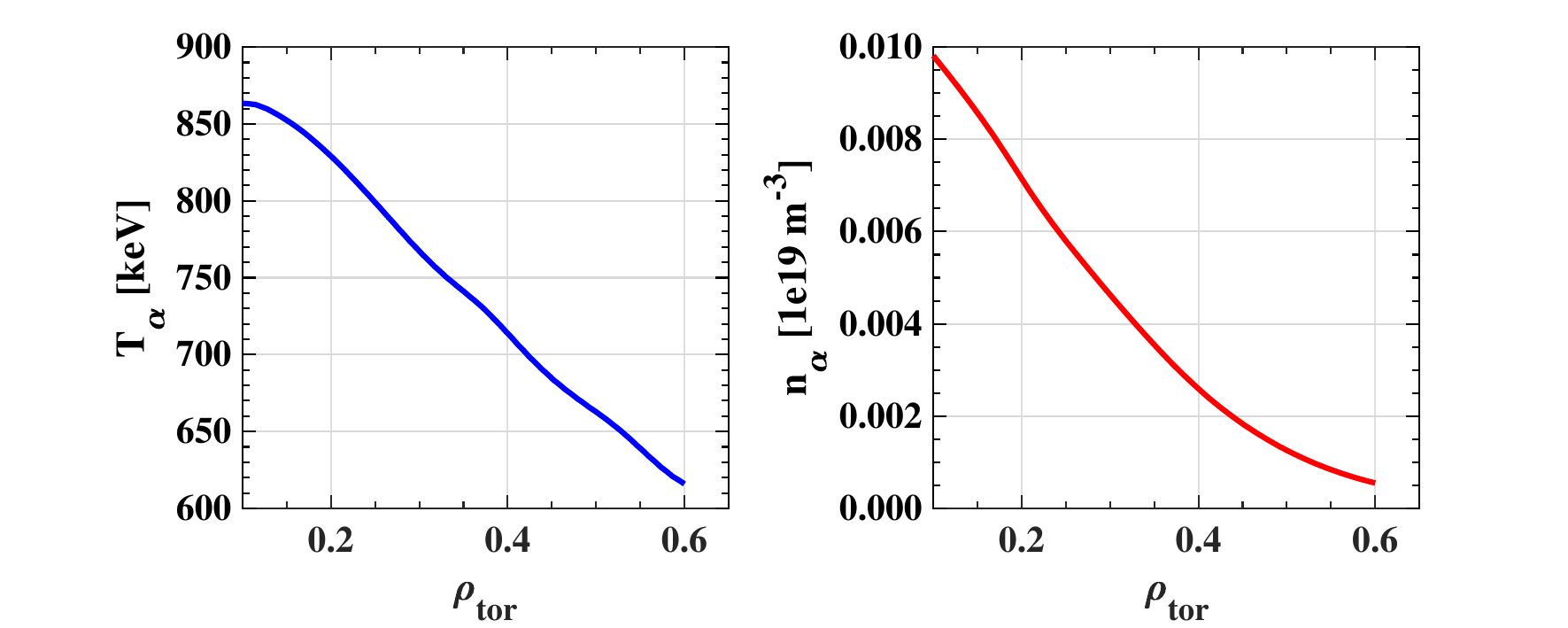}
\par\end{center}
\caption{Radial profile of the alpha particle (a) temperature and (b) density profiles computed by TRANSP for the JET Upgrade discharge $\#99912$ at $t = 8.6$s.}
\label{fig:fig_fast}
\end{figure}
The alpha particle temperature and density profiles computed by TRANSP \cite{transp_1,transp_2} are shown in Fig.~\ref{fig:fig_fast} for the radial domain covered by the GENE-Tango simulations, i.e., $\rho_{tor} = [0.1 - 0.6]$. These profiles are kept fixed within the GENE-Tango loop for the simulation retaining alpha particles in GENE. This stems from the significant contribution of alpha particles generated through beam-target reactions. While Tango can evolve self-consistently the alpha particle heating and profiles resulting from fusion reactions, it lacks a model to compute alpha particles generated by beam-target reactions.

\begin{figure*}
\begin{center}
\includegraphics[scale=0.35]{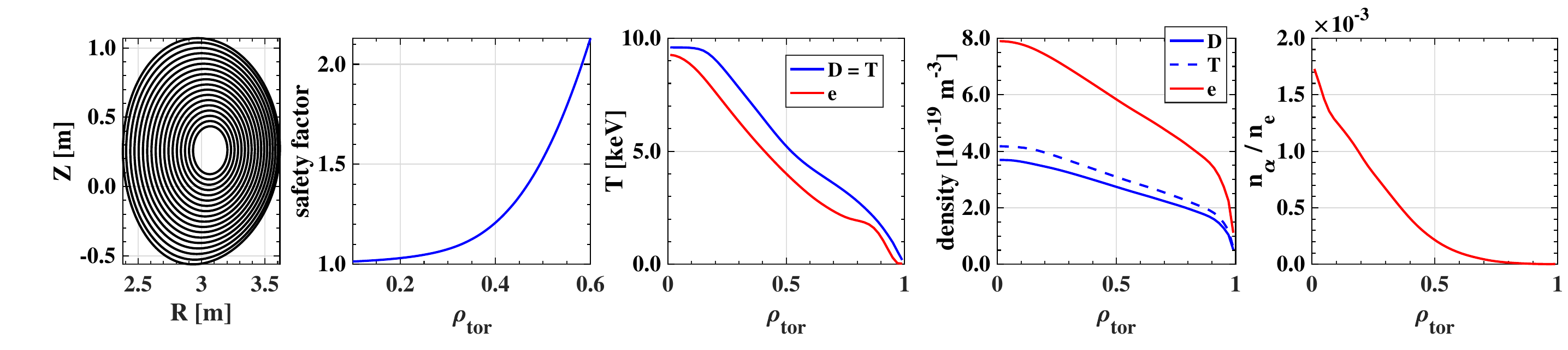}
\par\end{center}
\caption{a) Contours of constant poloidal flux of magnetic equilibrium reconstructed using pressure-constrained EFTP with TRANSP pressure profiles; (b) safety factor profile; (c) thermal deuterium, tritium and electron temperatures; d) corresponding density profiles of deuterium, tritium and electrons; e) ratio between alpha particle and electron density for the JET discharge $\#99912$ at $t = 8.6$s. The flux surfaces and safety factor profiles are shown within the simulated radial domain $\rho_{tor} = [0.1 - 0.6]$.}
\label{fig:fig_geo}
\end{figure*}

The magnetic equilibrium, reconstructed using pressure-constrained EFIT \cite{Szepesi_EPS_2021} with TRANSP pressure profiles including the full fast-ion pressure, is depicted in Fig.~\ref{fig:fig_geo} alongside the safety factor profile. Thermal ion (equal for deuterium and tritium) and electron temperature profiles are presented in Fig. ~\ref{fig:fig_geo}c, while deuterium, tritium, and electron density profiles are shown in Fig.~\ref{fig:fig_geo}d. These profiles are obtained from interpretative TRANSP simulations, fitting the experimental data points. In Fig.~\ref{fig:fig_geo}e, we provide the radial profile of the ratio $n_\alpha / n_e$, indicating that the alpha particle concentration remains below $0.15\%$, suggesting a negligible dilution effect at JET by alpha particles.

\begin{figure*}
\begin{center}
\includegraphics[scale=0.35]{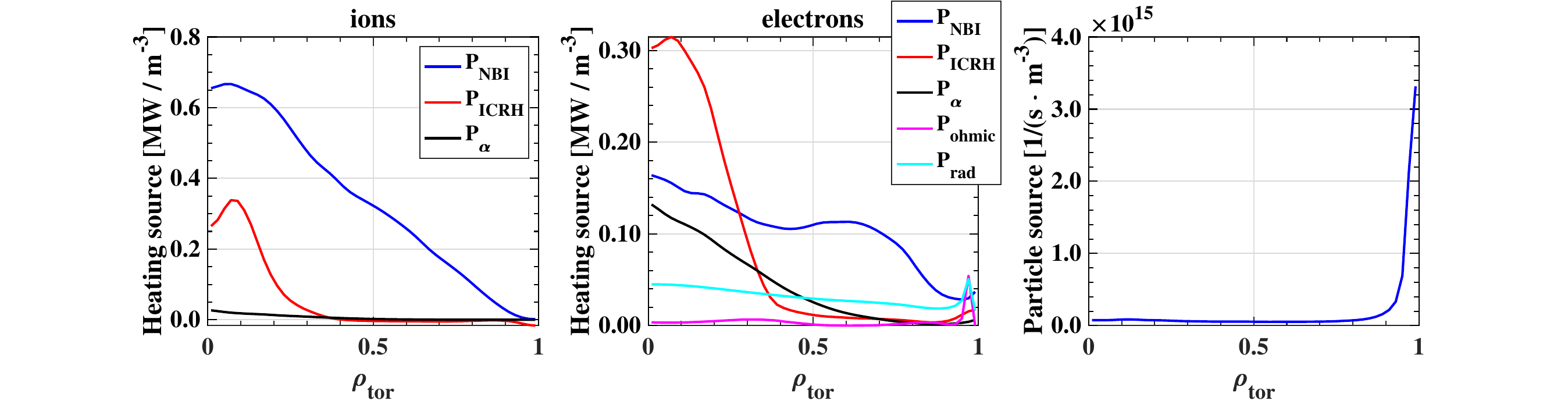}
\par\end{center}
\caption{Radial profiles of the ion and electron heating and particle sources for the JET discharge $\# 99912$ at $t = 8.6$s computed by TRANSP.}
\label{fig:sources}
\end{figure*}

Furthermore, we include in Fig.~\ref{fig:sources} the radial profiles of the heat and particle sources used in the GENE-Tango simulations. These profiles are extracted from TRANSP and remain fixed throughout the GENE-Tango loop. The only heat source permitted to adjust self-consistently at each GENE-Tango iteration is the collisional energy exchange between thermal ions and electrons. As shown in Fig.~\ref{fig:sources}, the alpha heating contribution is more substantial for electrons than for thermal ions, approximately four times larger on-axis. Additionally, we notice a non-negligible impact of the alpha heating on the electrons in comparison to their overall heating. The impact of the alpha heating on the electron temperature profile is investigated in detail in Section \ref{sec4}.

\section{GENE-Tango iterative framework} \label{sec2}

The numerical simulations summarized in this paper are performed using GENE-Tango, which integrates the gyrokinetic code GENE \cite{Jenko_PoP2000,Goerler_JCP2011} with the transport solver Tango \cite{Shestakov_JCP_2003}. While the coupling has been achieved with both the flux-tube \cite{Navarro_NF_2023} and radially global versions of GENE \cite{DiSiena_NF_2022,Disiena_NF}, this paper exclusively employs the radially global code. GENE-Tango allows for reliable profile predictions with large speed-ups compared to standalone flux-driven gyrokinetic simulations that require run times of the order of several confinement times. The approach aligns with the methodology used in reduced turbulence models and transport codes (e.g., TGLF-ASTRA \cite{Fable_PPCF_2013,Pereverzev_IPP_2002} and QuaLiKiz-JINTRAC \cite{Romanelli_PFR_2014}), where each code is simulating turbulence and transport phenomena solely on their inherent timescales \cite{Barnes_PoP_2010}. This is particularly efficient in the core of magnetic confinement plasma devices due to the large separation between drift-turbulence and collisional time scales. In this Section, we provide a brief overview of the iterative framework. For a more detailed description, we refer the readers to Ref.~\cite{Shestakov_JCP_2003,DiSiena_NF_2022}.

We start the iterative loop by running a stand-alone global GENE simulation, using as initial guess the profiles computed with TRANSP \cite{transp_1,transp_2}. This independent simulation continues until turbulent fluxes reach a stationary level. Afterwards, we average the fluxes (both heat and particle) over the saturated phase and transfer them to Tango. Tango solves the 1D transport equations for steady-state using together the GENE fluxes and the physical sources to evolve temperature and density profiles for each species. The new plasma profiles calculated by Tango are fed back into GENE, which restarts from the previous (saturated) simulation. It runs over a few eddy-turnover (correlation) times only and the new fluxes are transferred back to Tango. The process is repeated until turbulent fluxes align with the volume integral of injected heat and particle fluxes. As a final consistency check, we perform a stand-alone GENE simulation using the final plasma profiles computed by Tango. The coupled GENE-Tango simulation is considered fully converged only when this stand-alone GENE simulation yields the same fluxes as the last GENE-Tango iteration (matching the injected sources). 

It is worth mentioning that GENE applies Dirichlet boundary conditions, causing fluctuations to be damped inside the so-called buffer regions near each end of the radial domain. Consequently, within these buffer zones, it is not possible for the GENE fluxes to match the volume integral of the injected sources. This is particularly problematic for the outer buffer because of the large specific volume. Tango addresses this issue by implementing various strategies to extend turbulent fluxes inside the GENE buffer regions, mitigating undesired damping effects on plasma fluctuations. For the outer buffer, Tango interpolates the GENE fluxes in areas unaffected by the Krook operator. Beyond this region towards the outer boundary, turbulent fluxes are substituted with a linear extrapolation until reaching the end of the GENE-Tango grid. In the inner buffer, physical sources are deliberately set to zero, with minimal impact on the volume integral of the injected sources due to the reduced plasma volume near the magnetic axis. A detailed discussion on the Tango algorithm and the GENE-Tango coupling can be found in Refs.~\cite{Shestakov_JCP_2003, Parker_NF_2018, DiSiena_NF_2022}.

The use of GENE-Tango leads to a substantial reduction in computational expenses by several orders of magnitude compared with only gyrokinetic simulations, that require to be run for multiple confinement times to reach convergence, provided that transport and turbulence time scales are sufficiently decoupled. The reliability of GENE-Tango has been proved through several validation studies on numerous ASDEX Upgrade discharges, where the influence of energetic particles on plasma profiles was found to be negligible \cite{DiSiena_NF_2022}. Additionally, GENE-Tango has been recently validated also on high-performance scenarios at ASDEX Upgrade, where supra-thermal particle effects were essential to reproduce the experimentally observed peaking in the ion temperature profiles \cite{Disiena_NF}. An on-going effort is also focused on validating GENE-Tango in stellarator devices for turbulence-optimized configurations \cite{Navarro_NF_2023}, as well as across various W7-X experimental discharges \cite{Fernando_inpreparation}.

\section{Numerical setup and grid resolution} \label{sec3}

This paper presents radially global GENE-Tango simulations using a realistic ion-to-electron mass ratio, including collisions via a linearized Landau-Boltzmann collision operator conserving energy and momentum \cite{Crandall_CPC_2020}. The magnetic equilibrium remains fixed within the GENE-Tango loop to ensure consistency across the different cases analyzed in this paper, thus preventing discrepancies introduced by changes in the plasma geometry. The energetic particles generated via NBI and ICRF heating are not retained in the GENE simulations since they are considered to have only a minor effect on turbulent transport (having $\beta_{EP} / \beta_e \lesssim 0.3$). Here, GENE-Tango does not include the contribution of the neoclassical fluxes due to their negligible impact compared to the turbulent fluxes. For simplicity, each species is represented by a Maxwellian distribution function. Notably, the alpha particle species are modelled with an equivalent-Maxwellian distribution having the zeroth and second-order moments equal to the ones computed by TRANSP.

The plasma discharge $\#99912$ contains $47\%$ deuterium and $53\%$ tritium relative to the electron plasma density for the specified time slice. To speedup the computations, we have streamlined the simulations by considering only a single mixed DT ion species $(DT)$ with a relative mass of $m_{DT} = (n_D \cdot m_D + n_T \cdot m_T) / (n_D + n_T) =  1.265 m_D$, where $m_D$ and $m_{DT}$ denote the masses of deuterium and the mixed DT species, respectively. The choice of employing a mixed DT species rather than two distinct kinetic species is proved to have a minor effect on the GENE turbulent fluxes. This is addressed in Appendix \ref{sec9} on the final steady-state GENE-Tango profiles for the case retaining alpha particles. Similar results are also obtained when using TGLF-ASTRA.

These simulations performed in this paper use a grid resolution of $(n_x \times n_{k_y} \times n_z) = (384 \times 64 \times 32)$ in the radial $(x)$, bi-normal $(y)$, and field-aligned $(z)$ directions. The radial resolution corresponds roughly to two grid points per thermal ion-Larmor radius. Velocity space grids consist of $(n_{v_\shortparallel} \times n_\mu) = (48 \times 32)$ points, with $v_\shortparallel/v_{th,s} = [-3.5, 3.5]$ and $\mu B_0 / T_s = [0,12]$ for each species, with $B_0$ the on-axis magnetic field and $\mu$ the magnetic moment.

Here, $v_\shortparallel$ represents the velocity component parallel to the background magnetic field, and $\mu$ is the magnetic moment. In the radially global GENE-Tango simulations, the radial domain spans $\rho_{tor} = [0.1, 0.6]$, where $\rho_{tor}$ is the radial coordinate based on the toroidal flux $\Phi$, calculated as $\rho_{tor} = \sqrt{\Phi / \Phi_{LCFS}}$, with $\Phi_{LCFS}$ denoting the value of the toroidal flux at the last closed flux surface. The radial domain has been chosen to cover only $50\%$ of the overall radial domain to reduce the computational cost of gyrokinetic simulations while retaining the radial domain where alpha particle effects are expected to be more pronounced. Additionally, $v_\shortparallel$ is normalized in GENE to the thermal velocity of each species $s$, denoted as $v_{th,s} = \left(2T_s/m_s\right)^{1/2}$, where $T_s$ represents the temperature and $m_s$ stands for the mass. The magnetic moment is normalized to $B_0$, the on-axis magnetic field, and $T_s$. To alleviate the computational burden, only even toroidal modes (including the zonal component) are retained in the simulations. Therefore, $n_{0,min} = 1$ fishbones are not retained in the simulations. Moreover, fluctuations in $B_{\shortparallel}$ are neglected.

The global GENE simulations employ Krook-type particle and heat operators to maintain plasma profiles close to the reference profiles provided by Tango during each iteration \cite{Goerler_JCP2011}. This is achieved by applying heat ($\gamma_k$) and particle sources ($\gamma_p$) with amplitude set to $\gamma_k = 0.015 c_s / a$ and $\gamma_p = 0.15 c_s /a$, respectively. Here, $c_s$ represents the sound speed, calculated as $(T_e / m_i)^{1/2}$, with $T_e$ denoting the electron temperature at the reference radial position and $m_i$ representing the bulk ion mass in proton units. Additionally, we applied a numerical fourth-order hyperdiffusion method along the perpendicular dimensions and the field line to dampen fluctuations and mitigate the impact of unresolved scales and instabilities, such as electron temperature gradient (ETG) driven modes \cite{Pueschel_CPC_2010}. GENE enforces Dirichlet boundary conditions at both ends of the simulated radial domain using a Krook operator with an amplitude of $\gamma_b = 1.0 c_s/a$. This enforcement occurs within buffer regions, covering $5\%$ of the radial domain, where fluctuations are effectively suppressed to zero near the domain boundaries, thereby maintaining consistency with the Dirichlet boundary conditions. For simplicity, the toroidal rotation is not included in the GENE simulations.

Regarding the Tango numerical setup, we used relaxation coefficients for plasma pressure $(\alpha_p)$ and turbulent fluxes $(\alpha_q)$ to mitigate the strong dependence of the turbulent fluxes on the logarithmic gradient of the plasma temperature and density profiles and improve numerical stability \cite{Shestakov_JCP_2003, Parker_NF_2018, DiSiena_NF_2022}. In this paper, we set $\alpha_p = 0.1$ and $\alpha_q = 0.4$. Additionally, we run each GENE simulation within the GENE-Tango loop approximately for $t = 150 c_s/a$, which is enough for GENE solution to reach a quasi-stationary state.

\section{Alpha particle effects at JET with GENE-Tango} \label{sec4}

\begin{figure*}
\begin{center}
\includegraphics[scale=0.45]{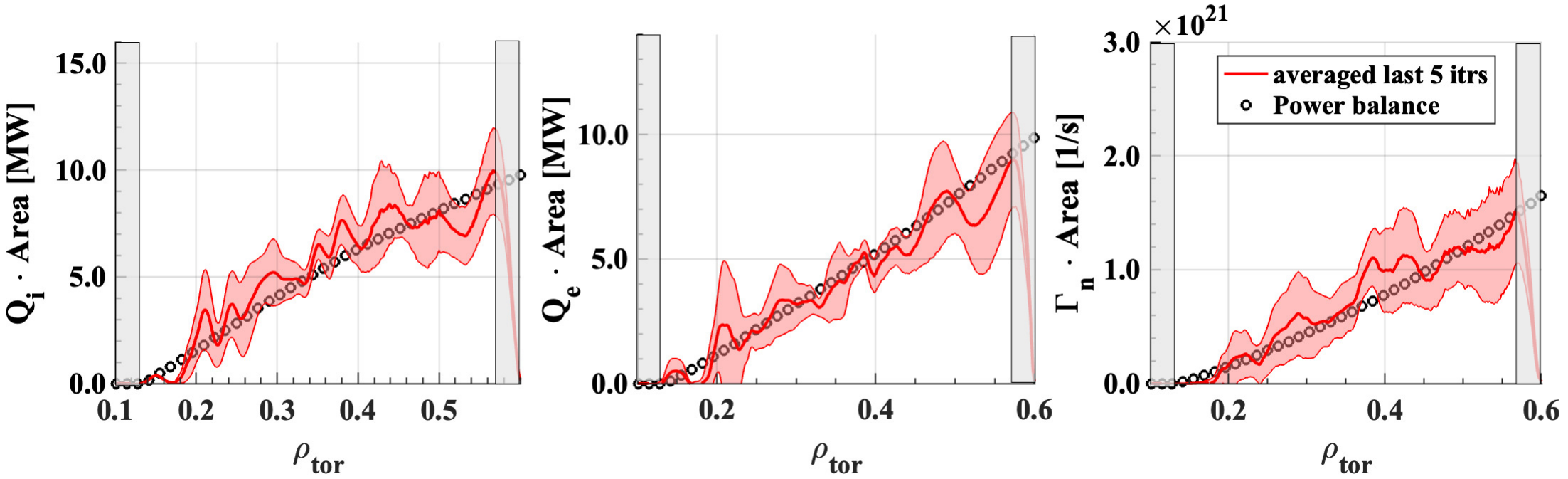}
\par\end{center}
\caption{Time-averaged radial profile of the (a) ion, (b) electron heat fluxes in MW and (c) particle flux in $1/s$ corresponding to the averaged last five GENE–Tango iterations (red) in the simulations retaining both alpha particles in GENE and alpha heating in Tango. The shaded gray areas denote the buffer regions and the black circles the volume integral of the injected particle and heat sources.}
\label{fig:fluxes}
\end{figure*}

\begin{figure*}
\begin{center}
\includegraphics[scale=0.45]{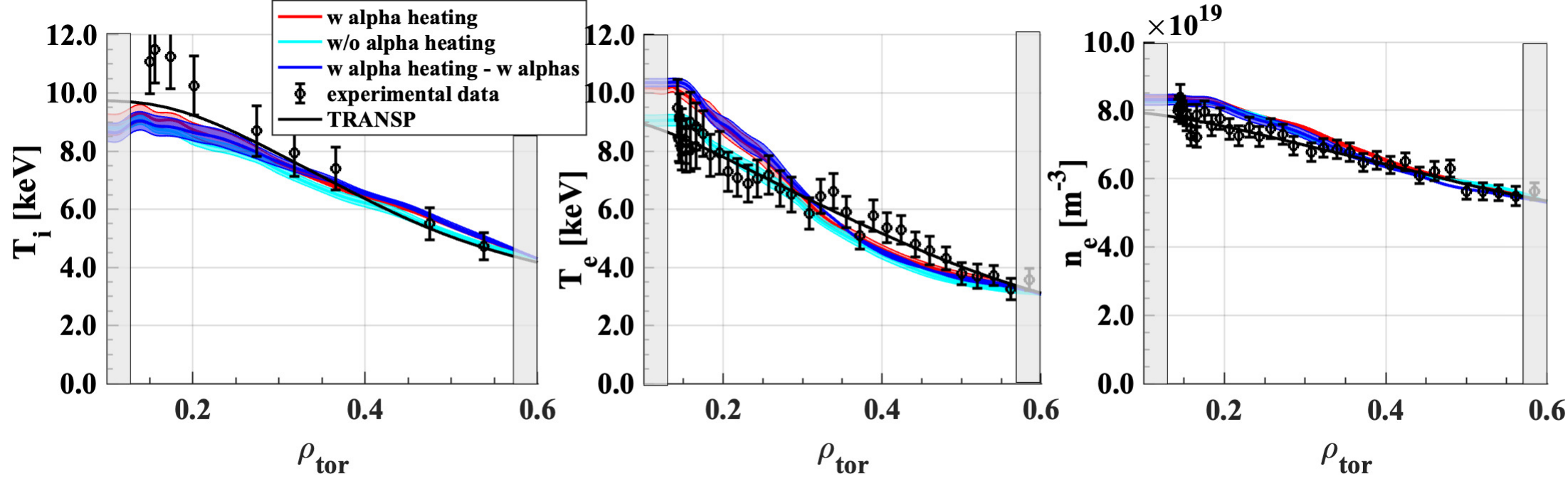}
\par\end{center}
\caption{Comparison of the (a) ion, (b) electron temperatures and (c) density computed by GENE–Tango for the different cases considered: with both alpha particles in GENE and alpha heating in Tango (red), without alpha particles in GENE but with alpha heating in Tango (blue), without alpha particles in GENE and alpha heating in Tango (cyan). Experimental measurements are represented by the black dots, while the initial TRANSP profiles are denoted by the black line. The TRANSP profiles are obtained from interpretative simulations fitting the experimental data points.}
\label{fig:fig_profiles}
\end{figure*}

This paper aims to explore the effect of alpha particles at JET during the D-T experimental campaign, specifically focusing on plasma discharge achieving the highest fusion power in 50-50 D-T in DTE2. This is studied by performing three different GENE-Tango simulations: i) retaining alpha heating in Tango while excluding alpha particles in GENE; ii) excluding both alpha heating and alpha particles in Tango and GENE, respectively; and iii) retaining both alpha heating and alpha particles in Tango and GENE. These simulations cover all the different cases to effectively asses the role of alpha particles at JET, considering their indirect effect as a heating source and direct effect on the underlying micro-turbulence. GENE-Tango simulations were performed employing the numerical setup described in Section \ref{sec3} for each of these cases. The thermal ion (mixed DT species) density is kept equal to the electron density profile for the simulations without alpha particle in GENE to ensure quasi-neutrality. The GENE-Tango simulations are run until the global turbulent fluxes computed by GENE match the volume integral of the injected sources at each location. This is illustrated, as an example, in Fig.~\ref{fig:fluxes} where the time-averaged GENE fluxes are presented for the last five GENE-Tango iterations. They are compared with the volume integral of the injected sources for each channel in the case where both alpha particles are retained in GENE and alpha heating is considered in Tango.
The final steady-state profiles are illustrated in Fig.~\ref{fig:fig_profiles} for each of the different cases and compared with the experimental measurements. The electron temperature and density measurements are obtained using Thomson Scattering, while the ion temperature is determined through Charge Exchange.
By looking at Fig.~\ref{fig:fig_profiles} we note a good agreement between the experimental profiles and the ion temperature, as well as the electron density profile computed by GENE-Tango. There are negligible variations observed across the different cases analyzed in this paper. The electron temperature is also well matched, showing minimal change whether we include or neglect the alpha particles in GENE. However, we observe a drop in the on-axis electron temperature when the alpha heating is removed from Tango. It reduces by approximately $1 keV$.
\begin{figure}
\begin{center}
\includegraphics[scale=0.40]{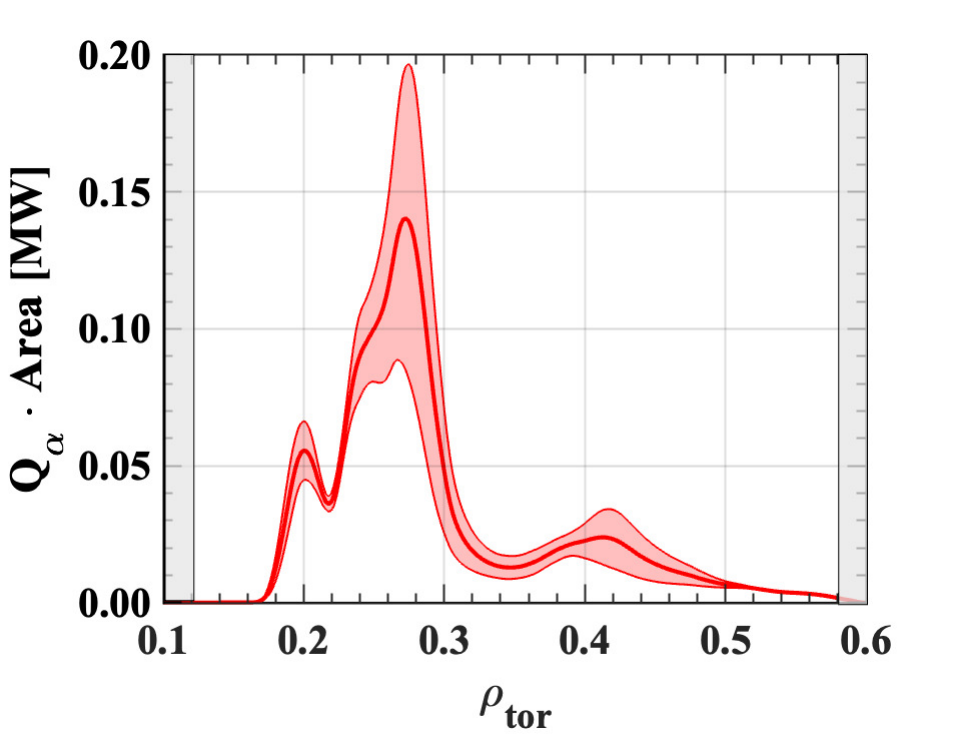}
\par\end{center}
\caption{Radial profile of the time-averaged alpha particle turbulent flux corresponding to the averaged last five GENE-Tango iterations in the simulation retaining both alpha particles in GENE and alpha heating in Tango.}
\label{fig:alpha_flux}
\end{figure}
Qualitatively, the profile agreements can be considered a successful validation of GENE-Tango at JET by closely replicating the experimental profiles across all the various channels but also highlight the visible influence of alpha heating on the electron temperature profile at JET. Interestingly, the impact of the alpha heating is found to be minimal on the ion temperature profile and plasma density. These results are consistent with the experimental observations in DT “afterglow” experiments at JET \cite{Kiptily_PRL_2023}. Additionally, the direct effect of alpha particles on turbulence is found to be negligible. This can also be inferred when looking at the alpha particle heat flux, as illustrated in Fig.~\ref{fig:alpha_flux}, where the alpha particle heat flux is smaller than the thermal species fluxes by several orders of magnitude. This finding is likely linked to the extremely low concentration of alpha particles in this particular plasma discharge, even amidst the large fusion output (as shown in Fig.~\ref{fig:fig_geo}e).

\section{Comparison with TGLF-ASTRA} \label{sec5}

We extend our studies of this JET discharge by comparing the GENE-Tango profiles, experimental measurements, and plasma profiles computed using TGLF-ASTRA. The TGLF-ASTRA simulations cover the same radial domain as GENE-Tango ($\rho_{tor} = [0.1 - 0.6]$) and employ the SAT2 saturation rule \cite{Staebler_NF_2021}. Similarly to our approach in the previous section, we compare the plasma profiles computed by retaining and neglecting the alpha heating contributions in ASTRA. We omit the case retaining alpha particles in TGLF and alpha heating in ASTRA due to the current limitations of TGLF in capturing some of the energetic particle effects on turbulence \cite{Doerk_NF_2017,Reisner_NF_2020,Mantica_PPCF_2019,Luda_NF_2021,Disiena_NF}. Additionally,  to facilitate a closer comparison with GENE-Tango, the TGLF simulations are performed enforcing quasi-neutrality, thus using $n_e = n_i$. These simulations used a single thermal ion species treated as a mixture of deuterium and tritium, neglecting impurities. However, impurities were taken into account for the calculation of collisional energy exchange in ASTRA. Simulations with D and T treated as separate kinetic species in TGLF give almost identical results to those obtained with only one lumped ion species.

\begin{figure*}
\begin{center}
\includegraphics[scale=0.45]{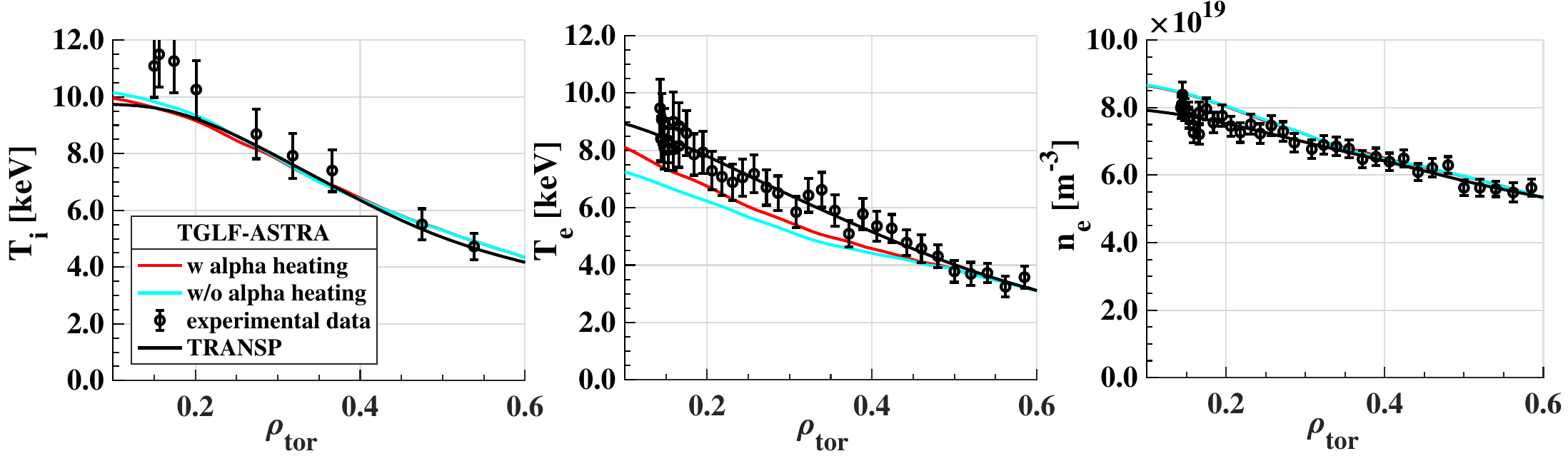}
\par\end{center}
\caption{Comparison of the (a) ion, (b) electron temperatures and (c) density computed by TGLF–ASTRA retaining (red) and neglecting (cyan) alpha heating in ASTRA. Experimental measurements are represented by the black dots, while the initial TRANSP profiles are denoted by the black line. The TRANSP profiles are obtained from interpretative simulations fitting the experimental data points.}
\label{fig:tglf}
\end{figure*}

The results of this comparison are summarized in Fig.~\ref{fig:tglf} showing the steady-state profiles.

We observe a good agreement for the plasma density and temperature profiles. Additionally, TGLF-ASTRA is consistent with the GENE-Tango results discussed in the previous sections. Specifically, while the influence of alpha heating on the ion temperature profile remains negligible in TGLF-ASTRA, a visible effect is observed on the electron temperature profile, which drops by 1keV drop on-axis when the alpha heating is neglected from the integrated modelling. These findings reinforce the capability of reduced turbulent models, like TGLF-ASTRA, to faithfully reproduce experimental measurements in high-performance JET discharges, consistently with the results of Ref.~\cite{Kim_NF_2023}. This is supported by the present study analyzing the JET discharge with the highest quasi-stationary peak fusion output throughout the entire JET DTE2 experimental campaigns in 50-50 D-T.

\section{Scan over alpha particle density} \label{sec6}

The previous sections have shown that, despite the minor but still visible increase in on-axis electron temperature profiles due to alpha heating at JET, the influence of alpha particles on micro-turbulence remains negligible. Here, we artificially rescale the alpha particle density computed by TRANSP to assess whether an increased alpha particle density can affect the turbulent fluxes of thermal species. This is done by designing four different cases. These cases include one represented by the alpha particle density computed by TRANSP and three others obtained by scaling up the alpha particle density with constant factors of 2.5, 5, and 10. In this way, we cover a broad spectrum of alpha particle densities, extending up to the expected alpha particle density of the ITER baseline scenario, approximately $n_\alpha / n_e \approx 1.2\%$ (corresponding to the case having the alpha density increased by a factor of 10) \cite{Bilato_PoP_2014}. 

\begin{figure}
\begin{center}
\includegraphics[scale=0.28]{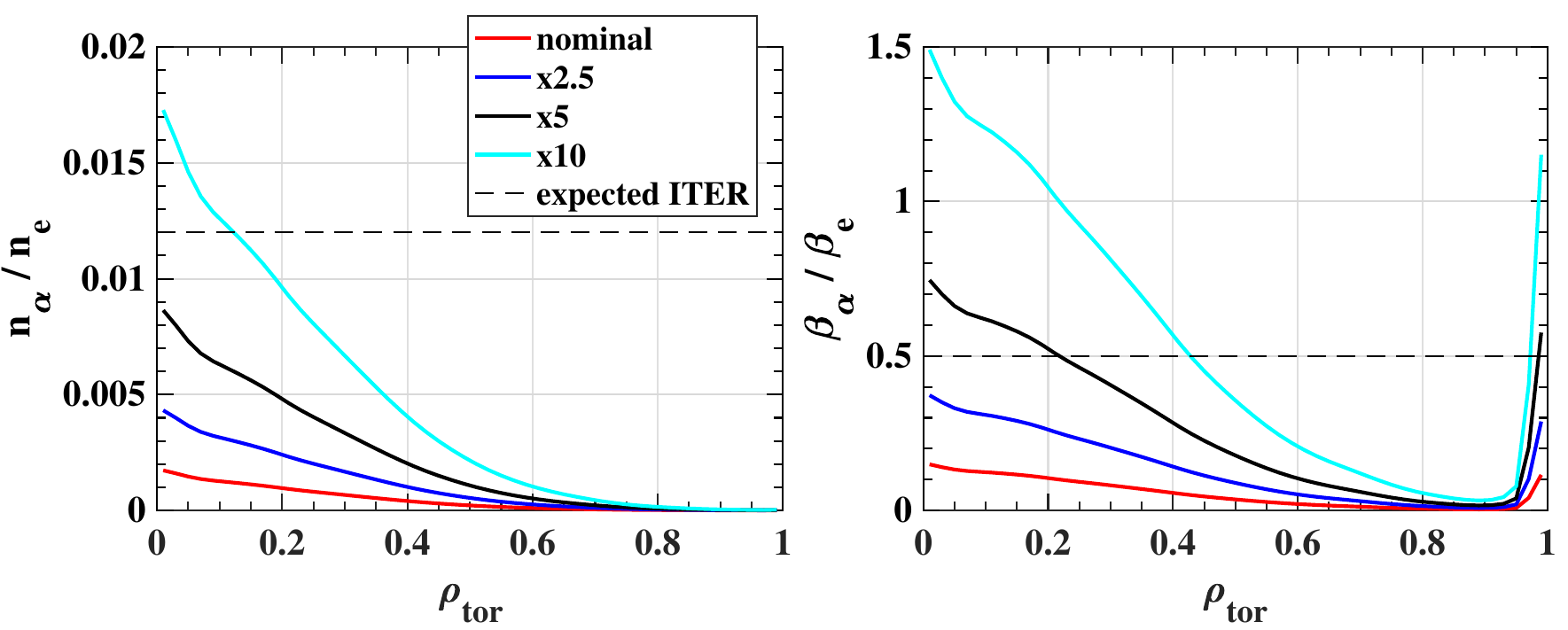}
\par\end{center}
\caption{Radial profile of the (a) alpha particle density profiles and (b) ratio between the alpha particle and electron plasma beta obtained from the nominal profile computed by TRANSP (red) and by rescaling the density by the constant factors 2.5 (blue), 5 (black), and 10 (cyan). The horizontal black line marks the on-axis value expected for ITER according to Ref.~\cite{Bilato_PoP_2014,Mantica_PPCF_2019}.}
\label{fig:scan_alpha_density}
\end{figure}

It is worth noting that the case with an alpha particle concentration of $n_\alpha / n_e \approx 1.2\%$ results in a $\beta_\alpha / \beta_e$ ratio that is roughly twice as large as that expected for ITER at $\rho_{tor} = 0.2$ \cite{Bilato_PoP_2014,Mantica_PPCF_2019} (see Fig.~\ref{fig:scan_alpha_density}), making direct extrapolation of these results to ITER conditions unfeasible. The thermal ion density is adjusted in each of these cases to maintain quasi-neutrality. The alpha particle concentrations relative to the electron density used for the simulations discussed in this section are illustrated in Fig.~\ref{fig:scan_alpha_density}.
It is important to note that the findings from this analysis cannot be directly extrapolated to ITER or other fusion reactors due to several limitations. Firstly, the plasma size, magnetic field, plasma current, and thermal pressure are not adjusted to ITER values but are kept fixed to the ones of the simulated JET experiment. Moreover, the simulations are performed using GENE stand-alone, rather than GENE-Tango, resulting in fixed plasma profiles that do not reach a proper steady-state and match the experimentally injected power. Despite these constraints, these analyses provide valuable insights into the potential impact of increased alpha particle density on background micro-turbulence.

\subsection{Linear AE stability analysis} \label{sec7}

We begin our investigation with linear stability analyses to identify potential high-frequency modes driven unstable by the increased alpha particle density. This analysis is performed by running linear global GENE simulations covering the same radial domain as the one used in the nonlinear GENE-Tango simulations, specifically $\rho_{tor} = [0.1 - 0.6]$. The plasma profiles employed correspond to those discussed in the previous section.

\begin{figure}
\begin{center}
\includegraphics[scale=0.28]{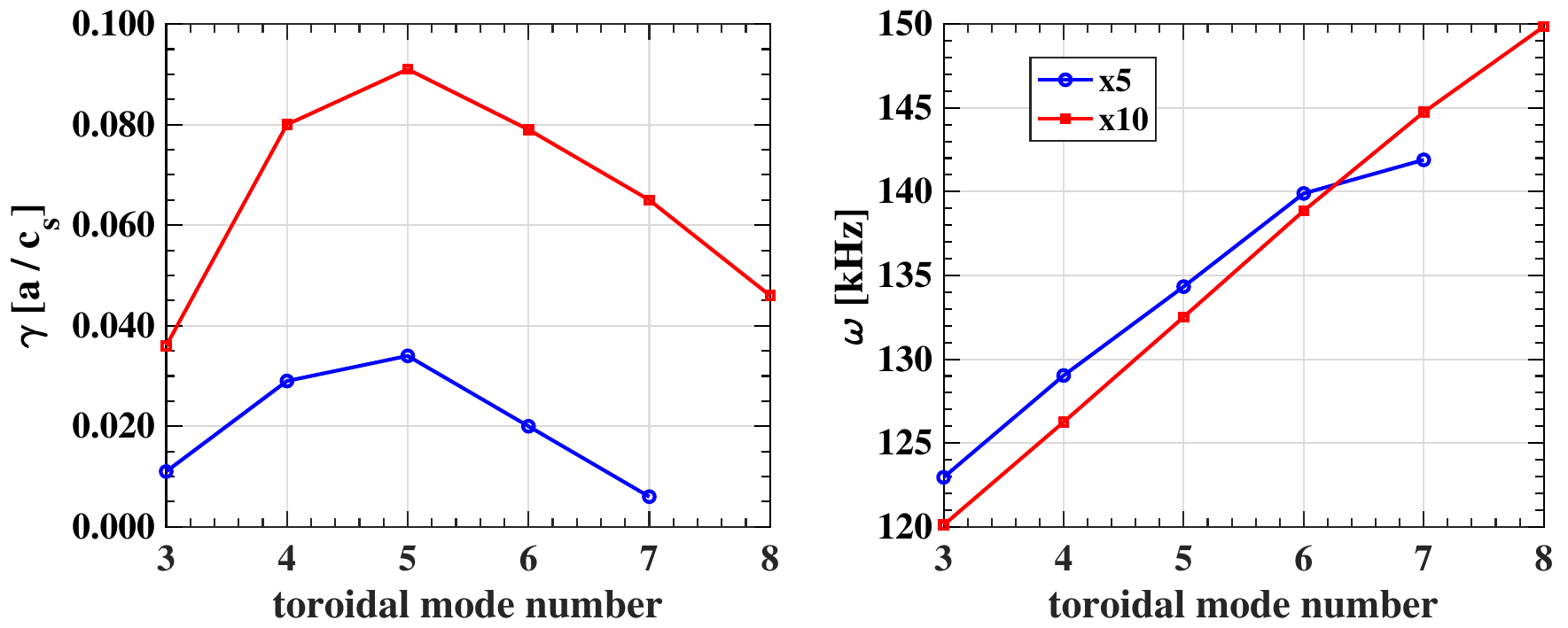}
\par\end{center}
\caption{Comparison of the linear growth rates (a) and frequencies (b) obtained by radially global linear GENE electromagnetic simulations for different toroidal mode numbers for the alpha particle density rescaled by a factor of five (blue) or ten (red) compared to the nominal density computed by TRANSP. The thermal profiles were initialized to those obtained from the GENE-Tango simulation retaining simultaneously alpha particles and alpha heating, with the only exception of the thermal ion density that was adjusted to fulfill quasi-neutrality.}
\label{fig:linear_growth_freq}
\end{figure}

\begin{figure}
\begin{center}
\includegraphics[scale=0.27]{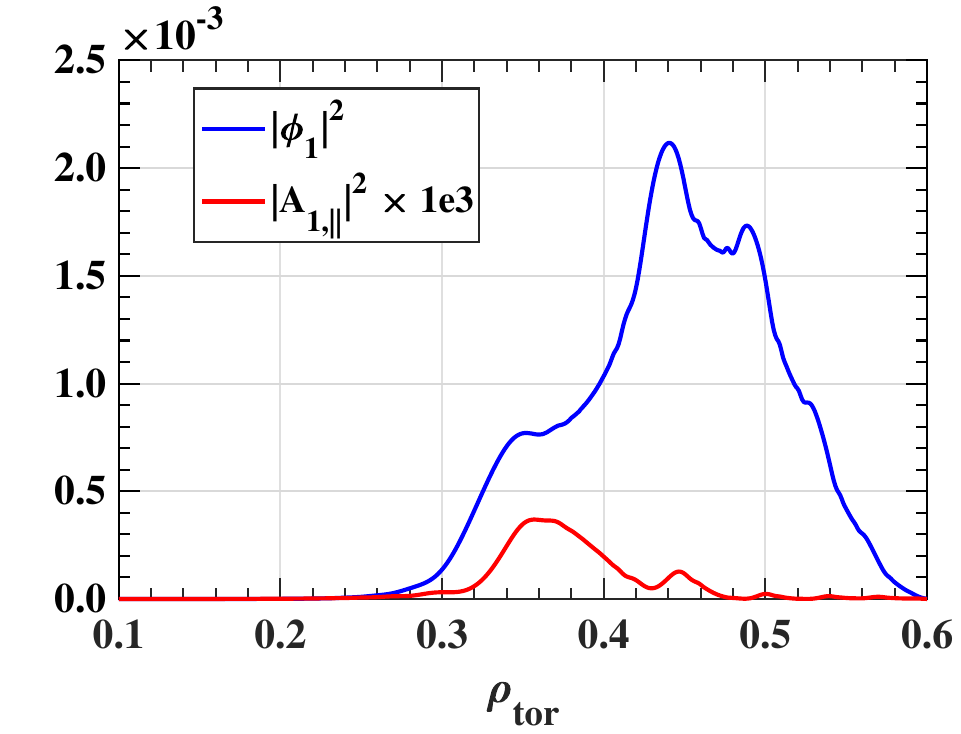}\includegraphics[scale=0.25]{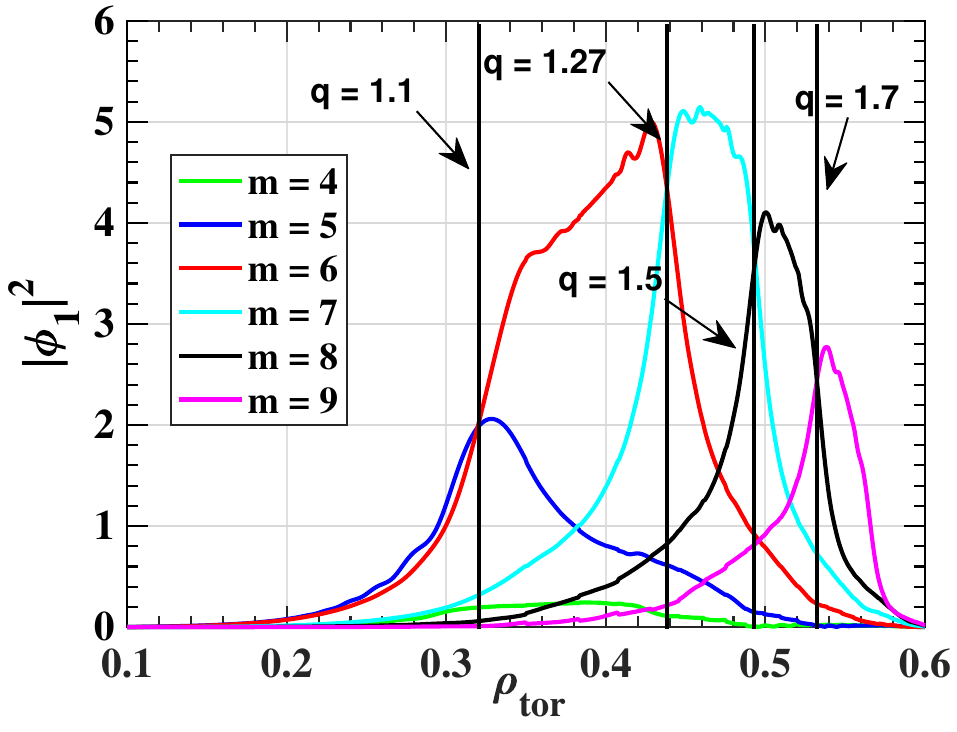}
\par\end{center}
\caption{Radial profile of the radial mode structure of (a) the (blue) electrostatic and (red) electromagnetic ($A_\shortparallel$) potentials and (b) poloidal harmonics for the most unstable toroidal mode number $n = 5$ in the electromagnetic GENE linear simulation with the alpha particle density increased by a factor of five compared to the nominal density computed by TRANSP. The vertical black lines denote the location of different rational surfaces.}
\label{fig:mode_struct}
\end{figure}

\begin{figure}
\begin{center}
\includegraphics[scale=0.40]{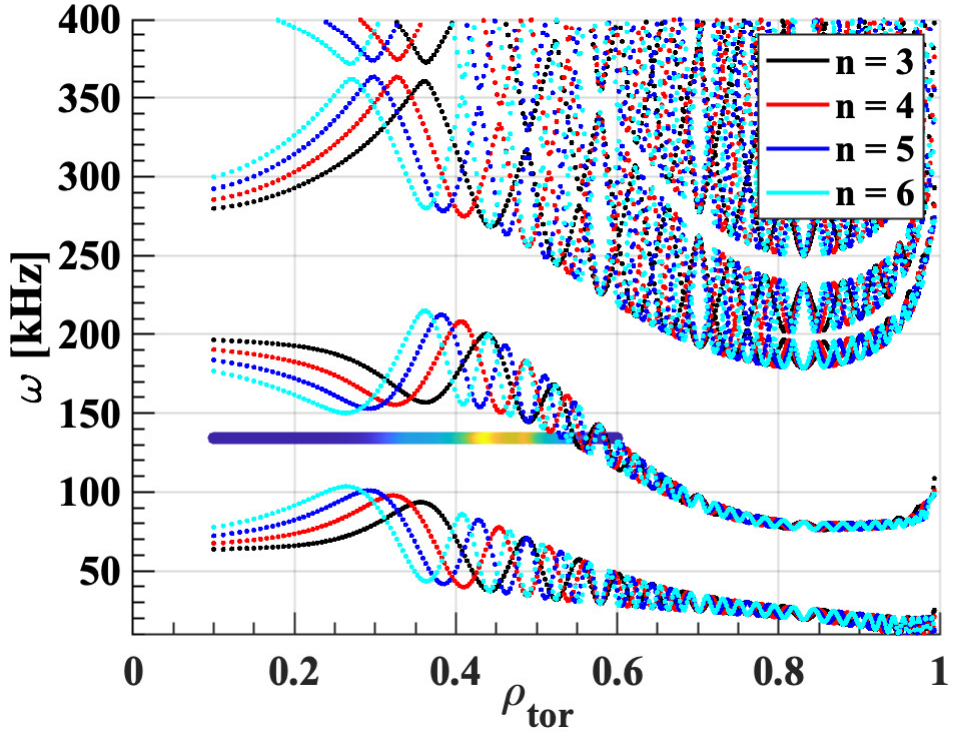}
\par\end{center}
\caption{Radial frequency spectrogram for the electrostatic potential for the toroidal mode number $n = 5$ obtained with linear GENE simulations. The thin dots represents the Alfv\'en continuum computed by the STELLGAP code considering the pressure up-shift BAE gap for different toroidal mode numbers.}
\label{fig:SAW}
\end{figure}

The results of the GENE simulations are shown in Fig.~\ref{fig:linear_growth_freq}, illustrating the most unstable linear growth rates and frequencies across various toroidal mode numbers. While no unstable alpha particle-driven modes had been found at nominal and by a factor of 2.5 increased alpha density, they are present at a factor of 5 increase compared to the one computed by TRANSP. The most unstable mode is $n = 5$. Fig.~\ref{fig:linear_growth_freq} does not show the linear growth rates and frequencies for both the nominal alpha particle density and the density increased by a factor of 2.5 since no high-frequency modes are found unstable under these conditions.
The mode structure of the most unstable alpha-particle driven mode ($n = 5$) is illustrated in Fig.~\ref{fig:mode_struct} for the case having a fivefold increase in the alpha particle density compared to the one computed by TRANSP. Notably, this mode exhibits a radially broad structure along the radial direction, with the electromagnetic component being negligible in comparison to its electrostatic counterpart. Fig.~\ref{fig:mode_struct} also contains the radial profiles of poloidal harmonics of the electrostatic potential for toroidal mode number $n = 5$. Fig.~\ref{fig:mode_struct} illustrates that the dominant poloidal mode numbers associated with toroidal mode $n = 5$ are $m = 6$, $m = 7$, and $m = 8$, showing a beating pattern in the poloidal harmonics. This is consistent with similar findings observed with ORB5 \cite{Lanti_CPC_2020, Mishchenko_CPC_2019} and LIGKA ~\cite{Lauber_JCP_2007} when performing linear stability analyses for the ITER standard scenario. Additionally, Fig.~\ref{fig:mode_struct} provides the local values of the safety factor profile corresponding to the locations of the various poloidal harmonics.
To elucidate the nature of these high-frequency modes, Fig.~\ref{fig:SAW} shows the Alfv\'en continuum obtained using the STELLGAP code \cite{Spong_PoP_2003}, incorporating the pressure up-shift BAE gap \cite{Lauber_PR_2013}. Additionally, on the same figure we showcase the mode structure of the most unstable mode at $n = 5$, using an alpha particle density five times higher than the nominal value computed by TRANSP. Fig.\ref{fig:SAW} demonstrates that the high-frequency modes observed in simulations with increased alpha particle density are unstable Toroidal Alfv\'en Eigenmodes (TAE modes) destabilized within the SAW continuum gaps \cite{Heidbrink_PoP_2008,Gorelenkov_NF_2014,Chen_RMP_2016}. 

\subsection{Nonlinear simulations} \label{sec8}

After performing linear stability analyses in the previous section, we found that alpha particles destabilize TAE modes when their density is increased above five times the nominal value computed by TRANSP. In this section, we perform stand-alone nonlinear global GENE simulations for each of the cases discussed in Fig.~\ref{fig:scan_alpha_density}. The numerical setup and grid resolutions are the same as those outlined in Section \ref{sec3} for the simulation where the alpha particle density is set to 2.5 times the nominal density. For the other simulations, we have adjusted the minimum finite toroidal mode number to $n_{0,min} = 1$ and increased the resolution along the bi-normal direction to $n_{k_y} = 96$. This modification became necessary to resolve the turbulent spectra correctly, which shifted towards lower toroidal mode numbers in simulations with an increased alpha particle density. Additionally, we employed radially-dependent block-structured grids to better resolve the velocity grids without increasing the resolution along these directions \cite{Jarema_CPC_2016}.

\begin{figure*}
\begin{center}
\includegraphics[scale=0.40]{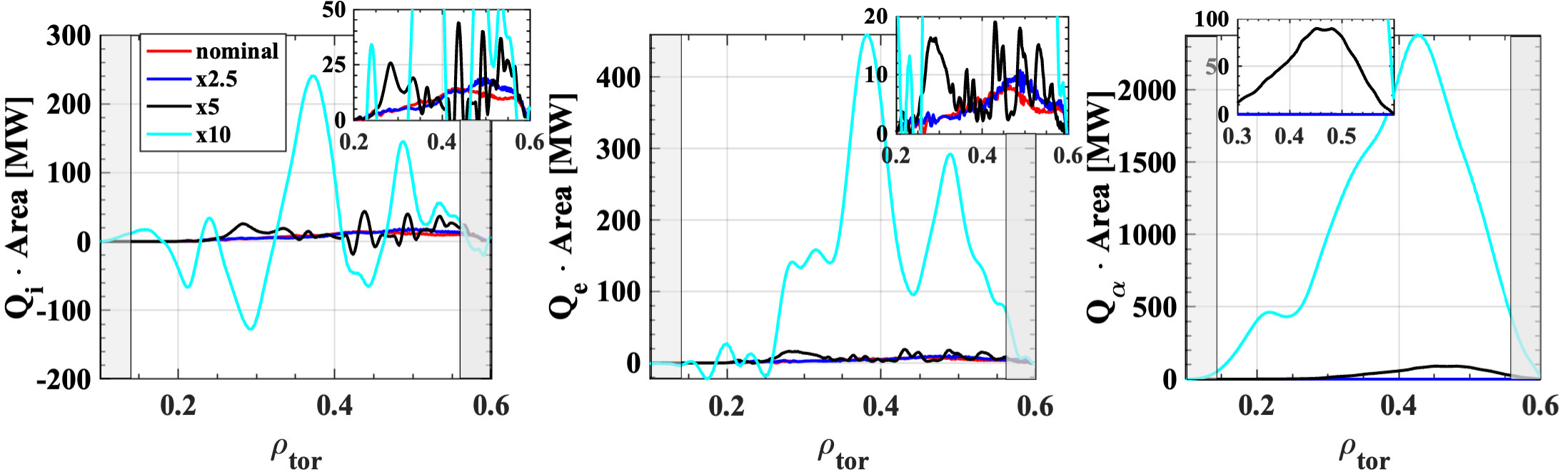}
\par\end{center}
\caption{Time-averaged radial profile of the (a) ion, (b) electron and (c) alpha particle heat fluxes in MW obtained in the global GENE simulations over the saturated phase using the different alpha particle density profiles illustrated in Fig.~\ref{fig:scan_alpha_density}. The inlays contains a zoom into the turbulent fluxes of each species in $\rho_{tor} = [0.2 - 0.6]$.}
\label{fig:scan_alpha_flux}
\end{figure*}

\begin{figure}
\begin{center}
\includegraphics[scale=0.40]{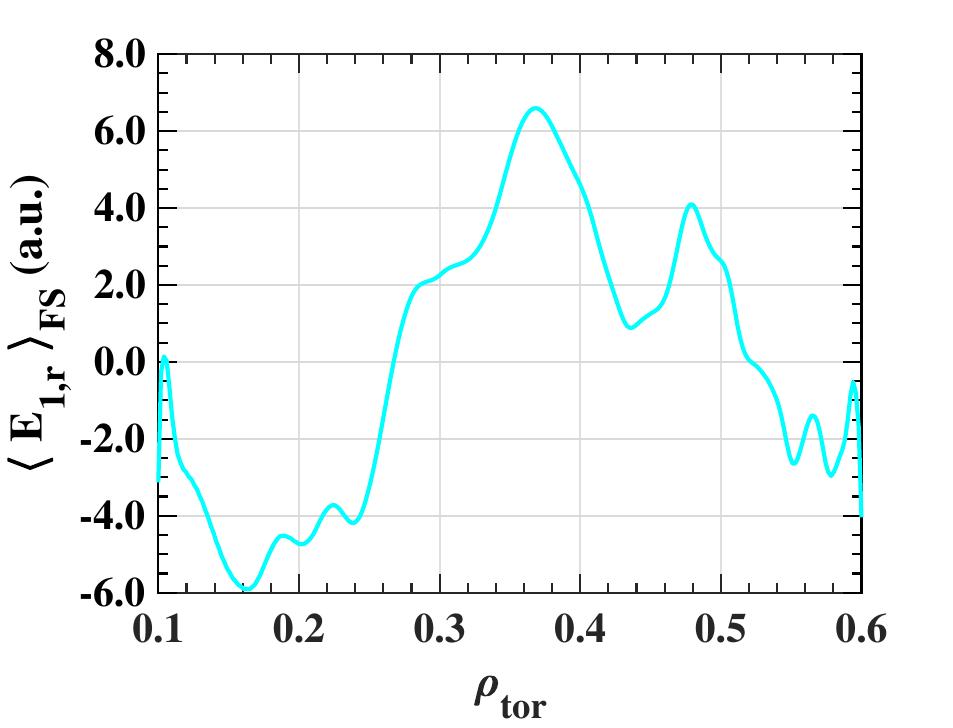}
\par\end{center}
\caption{Radial profile of the normalized flux-surface averaged radial electric field $\langle E_{1,r} \rangle_{FS} = - \partial_{\rho_{tor}} \phi_1(n=0)$ over the saturated phase obtained in the nonlinear global GENE simulation using the alpha particle density increased by a factor of ten compared to the nominal one computed by TRANSP.}
\label{fig:Er}
\end{figure}

The resulting turbulent fluxes for the thermal (DT) ions, electrons, and alpha particles are depicted in Fig.~\ref{fig:scan_alpha_flux}.
While negligible differences are noted when the alpha particle density is increased by a constant factor of 2.5 compared to the nominal density computed by TRANSP, turbulent fluxes undergo a visible increase when the alpha particle density is further increased by a factor of 5 in all the different channels. An oscillatory pattern develops for the thermal ion and electron turbulent fluxes corresponding to the different poloidal harmonics of the unstable alpha particle driven TAE. The alpha particle heat flux increase up to large fluxes reaching $Q_\alpha$ = 90 MW at $\rho_{tor} = 0.45$. When the alpha particle density is increased by a factor of 10 compared to the value computed by TRANSP, all turbulence fluxes exhibit a large destabilization. Interestingly, the alpha particle heat flux experiences a substantial four-order-of-magnitude increase, leading to a significant rise in the electron heat flux as well. Moreover, the oscillatory pattern of the thermal ion heat flux becomes more pronounced with increasing alpha particle density. This oscillatory structure along the radial direction for thermal ion heat flux mirrors the behavior of the flux-surface averaged radial electric field, driven by unstable TAE modes. This is depicted in Fig.~\ref{fig:Er}. While the thermal ion and alpha particle heat fluxes are predominantly electrostatic, the electron flux is dominated by the electromagnetic (flutter) component. These findings align with prior studies indicating that unstable fast ion modes contribute significantly to the enhanced electromagnetic flux of electrons \cite{Citrin_PPCF_2015, Garcia_NF_2015, DiSiena_JPP_2021}.

\begin{figure*}
\begin{center}
\includegraphics[scale=0.40]{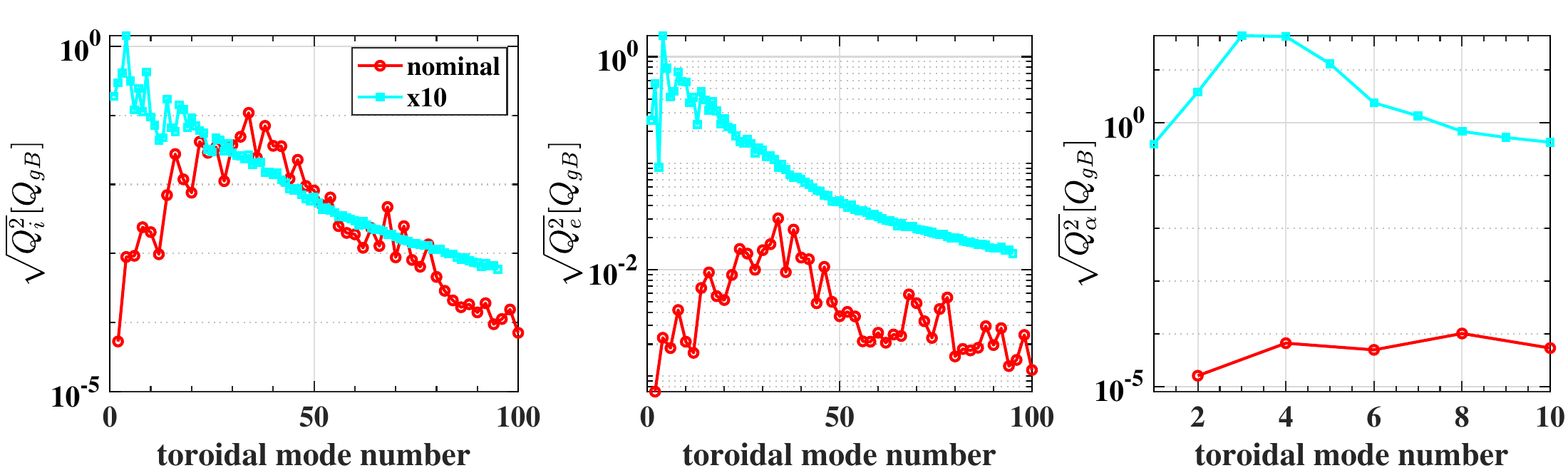}
\par\end{center}
\caption{Comparison of the nonlinear heat flux spectra (a) for the thermal ions, (b) electrons and (c) alpha particles obtained in the electromagnetic GENE simulation using the nominal (red) and the ten-time enhanced (cyan) alpha particle densities for different toroidal mode numbers. The logarithmic scale is required to make the spectra for the nominal alpha particle density visible.}
\label{fig:nonlinear_spectra}
\end{figure*}

The heat flux spectra exhibit large modifications when the alpha particle density is increased by a factor of 10. This is shown in Fig.~\ref{fig:nonlinear_spectra}, illustrating the time-averaged heat flux spectra during the stationary phase, obtained as the sum of the electrostatic and the electromagnetic components. The spectra is also averaged along the radial direction for each species considered in the GENE simulations. Fig.~\ref{fig:nonlinear_spectra} reveals a visible shift in turbulent heat flux spectra towards smaller toroidal mode numbers when the alpha particle density is increased by factors of 5 and 10, showing that most of the entire turbulent fluxes are predominantly driven by low toroidal mode numbers $n = [1 - 10]$. Furthermore, we note a large flux increase for both thermal ions and electrons at the toroidal mode number $n = 3$, and exclusively for thermal ions at its harmonics $n = 6$. This observation hints at a nonlinear interaction with the unstable TAE modes.

The presence of these TAE modes induces a significant increase in the overall turbulent fluxes, as illustrated in Fig.~\ref{fig:scan_alpha_flux}, pushing the fluxes further away from the experimental power balance. This effect is particularly pronounced for electrons. The increase in the turbulent fluxes in the presence of unstable fast ion driven modes is observed in both local \cite{Garcia_PPCF_2022,Citrin_PPCF_2023}, global gyrokinetic simulations \cite{DiSiena_NF_2023_2,Disiena_NF,Biancalani_PPCF_2021,Ishizawa_NF_2021} and experiments \cite{Stutman_PRL_2009,Ren_NF_2017}. With the increase in electron heat flux, a GENE-Tango simulation would evolve the plasma profiles by flattening the gradients of electron pressure to mitigate the large fluxes. Consequently, this would lead to a reduction in plasma beta and the drive of TAE modes. Understanding this delicate balance between fast ion-driven modes and turbulence requires simulations capable of self-consistently evolving plasma profiles and turbulence. Such simulations have been conducted for an ASDEX Upgrade discharge \cite{Disiena_NF} and future studies will be performed with GENE-Tango for burning plasma experiments retaining alpha particles into the modelling.

Moreover, we observe that the structure of the electrostatic potential exhibits large-scale features in configuration space when alpha particle-driven TAE modes are unstable. This observation aligns well with recent GKNET findings, which indicate homogenization across the poloidal cross-section of turbulence, characterized by large-scale structures \cite{Ishizawa_NF_2021}. This homogenization amplifies fluctuation levels and turbulent transports, ultimately deteriorating fusion performance.
\begin{figure}
\begin{center}
\includegraphics[scale=0.37]{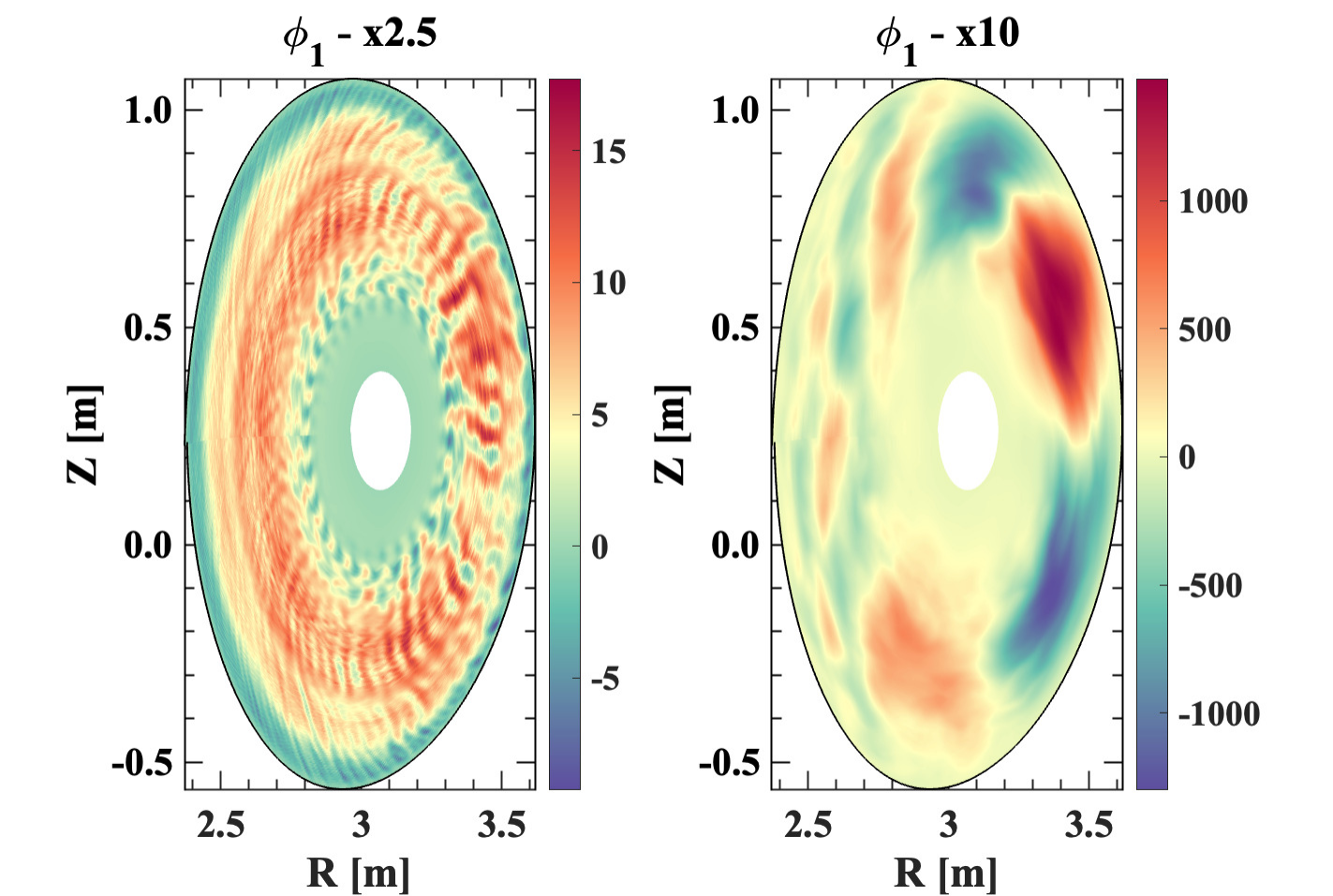}
\par\end{center}
\caption{Snapshots of the poloidal cross section of the electrostatic potential for the nonlinear GENE global simulations using an increased alpha particle density by a factor of (a) 2.5 and (b) 10 compared to the nominal density computed by TRANSP. The continuous black line delimits the boundary of the flux surface corresponding to the radial location $\rho_{tor} = 0.6$, representing the last radial grid point considered in the GENE global simulations.}
\label{fig:torus}
\end{figure}

However, there are several caveats to consider regarding the results discussed in this section. Firstly, the destabilization of a global TAE mode by alpha particles, along with the beating of different poloidal harmonics — particularly pronounced when the alpha particle density was increased tenfold compared to the nominal value computed by TRANSP — suggests the possible need for an expanded radial domain in this specific case. However, performing a nonlinear simulation with an expanded radial domain was not feasible, as the current simulation already required approximately 15 million CPU hours. Additionally, it is worth noting that the simulations in this section employed fixed profiles for thermal species, obtained from steady-state GENE-Tango simulations with alpha particles and alpha heating. The only exception was the thermal ion density, which was adjusted to maintain quasi-neutrality for each alpha particle density. Consequently, the increased alpha particle density cannot be self-consistently sustained by the thermal species pressure. Despite these limitations, the results suggest that alpha particle-driven TAEs can indeed influence background turbulent fluxes. These preliminary findings indicate a potentially detrimental impact, consistent with results from other studies under different plasma conditions \cite{Ishizawa_NF_2021}.

\section{Conclusions} \label{sec9}

This paper analyzes the impact of alpha particles on plasma performances at JET for the plasma discharge $\#99912$, achieving the highest quasi-stationary peak fusion performance in the DTE2 experimental campaigns in 50-50 D-T. This involves analyzing both the direct influence of alpha particles on plasma micro-turbulence and the role of alpha heating on plasma profiles. The simulations are performed using the radially global version of the gyrokinetic code GENE coupled to the transport solver Tango. This code coupling enables reliable profile predictions exploiting the large time-scale separation between microscopic and macroscopic physics, reducing the required computational cost by several orders of magnitude compared to stand-alone flux-driven simulations. 

The role of alpha particles at JET is investigated by performing GENE-Tango simulations for three different plasma setups: i) retaining alpha heating in Tango while excluding alpha particles in GENE; ii) excluding both alpha heating and alpha particles in Tango and GENE, respectively; and iii) retaining both alpha heating and alpha particles in Tango and GENE. The numerical profiles computed by GENE-Tango are compared against experimental measurements, demonstrating excellent agreement in each channel (ion, electron temperatures and plasma density). Our analysis reveals that alpha particles have a negligible influence on turbulent transport in this JET discharge, with GENE-Tango converging to similar plasma profiles regardless of including alpha particles as kinetic species in GENE. However, alpha heating contributes to the peaking of the electron temperature profile, increasing by 1 keV in the on-axis temperature when alpha heating is included in Tango. The limited impact of alpha particles on turbulent transport observed in this JET discharge - despite this being the plasma discharge with the highest fusion output in DTE2 - is attributed to the low concentration of alpha particles.

Similar results are also observed when employing TGLF-ASTRA with the SAT2 saturation rule to model this JET discharge. Specifically, we observe that TGLF-ASTRA not only consistently matches the experimental measurements, albeit with a slight underestimation in electron temperature, but also reproduces the 1 keV reduction in on-axis electron temperature seen in GENE-Tango when alpha heating is neglected in ASTRA. These findings demonstrate the capability of TGLF-ASTRA to capture key features of high-performance JET discharges.

Additionally, we designed four different plasmas to evaluate alpha particles' potential impact on plasma micro-turbulence under conditions with larger alpha particle density. In these setups, the alpha particle concentration was artificially increased by a constant factor from the reference profile computed by TRANSP up to values expected for ITER. The thermal profiles were kept fixed to those calculated by GENE-Tango using the nominal alpha particle density, except the thermal ion density, which was adjusted to ensure quasi-neutrality. We performed linear and nonlinear GENE standalone simulations for each of these setups.

Our findings revealed that beyond a fivefold increase in the alpha particle density, alpha particle-driven TAEs were destabilized, with their growth rates scaling with the alpha particle density. Interestingly, in the setup featuring unstable TAEs and an alpha particle density five times larger than the nominal one computed by TRANSP, we observed a large enhancement in the alpha particle turbulent fluxes, leading to a progressive increase in the thermal ions and electron fluxes. Furthermore, when in these simulations of JET plasmas the alpha particle density is scaled up to the values expected for ITER, namely tenfold that computed by TRANSP, alpha particle turbulent fluxes reach up to 2500 MW. This pronounced increase in the alpha particle fluxes, while unsustainable in a plasma reactor, induced oscillatory patterns in thermal ion turbulent fluxes, characterized by regions of large turbulent fluxes and areas with negative flux. This was attributed to the generation of substantial radial electric fields that violently sheared apart the large turbulent eddies. However, regardless of this radial electric field, we observe an overall notable confinement degradation.

Although these findings cannot be directly extrapolated to ITER since the relevant plasma parameters (both dimensional and dimensionless) are different between these JET plasma simulations and the ITER conditions, they do suggest that alpha particles can have a noticeable impact on performance. These results emphasize the importance of future modelling studies, especially those focused on fusion reactors, to include alpha particles in their simulations to thoroughly assess the performance of upcoming fusion reactors.

\appendix
\section{Impact of using a single DT-mixture ion species} \label{sec10}

\begin{figure*}
\begin{center}
\includegraphics[scale=0.40]{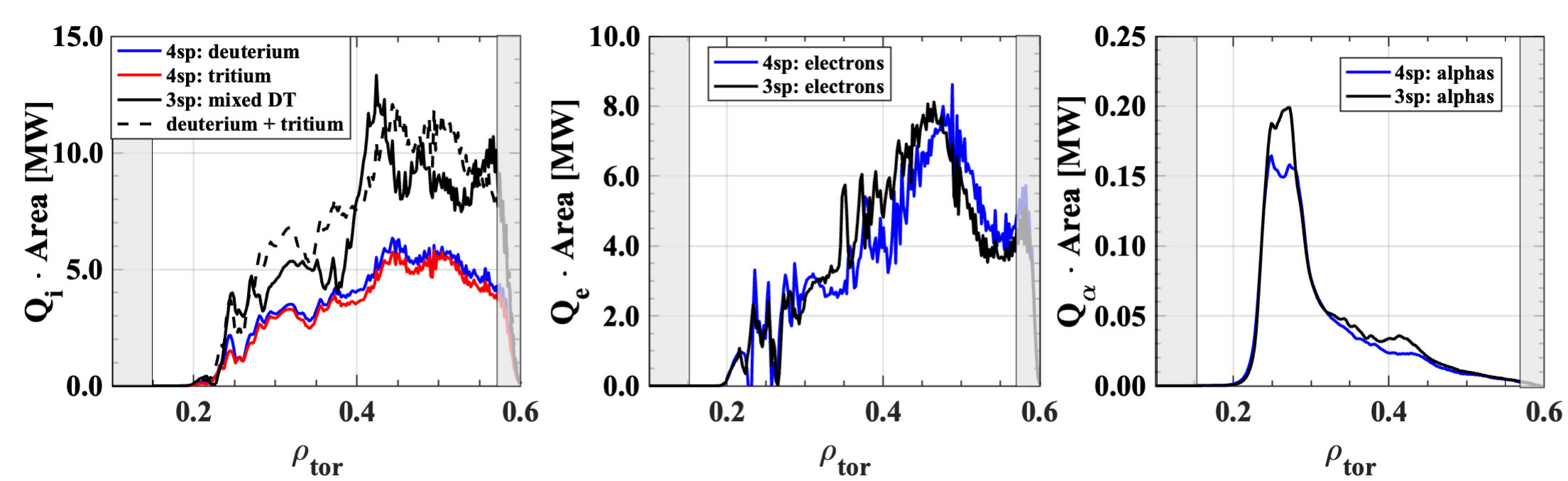}
\par\end{center}
\caption{Contour plots of growth rates (a) and frequencies (b) obtained by flux-tube simulations at different radial locations for different values of the toroidal mode number using the steady-state GENE-Tango profiles.}
\label{fig:appendix}
\end{figure*}

Below we present evidence that using a single ion species, represented by a mixture of deuterium and tritium, rather than treating them as separate species, does not yield visible differences in the total ion turbulent fluxes. This is depicted in Fig.~\ref{fig:appendix}, where the radial profiles of the turbulent fluxes for each species are compared between the simulation using deuterium and tritium and the one utilizing a mixed DT species.
These findings are noteworthy and enable a reduction in the number of simulated species in the gyrokinetic simulations, leading to a significant decrease in computational time. This becomes particularly crucial when considering future simulations of ITER, where the grid resolution and computational time for such simulations are exceptionally expensive. The ability to streamline the number of simulated species holds the potential to considerably accelerate gyrokinetic simulations in these demanding scenarios.

\section*{Acknowledgements}

The authors would like to acknowledge insightful discussions with P. Mantica, E. Fable, G. Tardini, and C. Bourdelle. This work has been carried out within the framework of the EUROfusion Consortium, funded by the European Union via the Euratom Research and Training Programme (Grant Agreement No 101052200 — EUROfusion). Views and opinions expressed are however those of the author(s) only and do not necessarily reflect those of the European Union or the European Commission. Neither the European Union nor the European Commission can be held responsible for them. Numerical simulations were performed at the Marconi, Leonardo and JFRS Fusion supercomputers at CINECA, Italy.

\end{document}